\documentclass{JHEP3}    
\usepackage{epsfig}
\usepackage{subfigure}
\usepackage{psfrag}
\usepackage{graphicx}

\newcommand{\eref}[1]{(\ref{#1})}

\newcommand{\fref}[1]{Figure~\ref{#1}}
\newcommand{\cref}[1]{Chapter~\ref{#1}}
\newcommand{\beq}{\begin{equation}}
\newcommand{\eeq}{\end{equation}}
\newcommand{\ba}{\begin{array}}
\newcommand{\ea}{\end{array}}
\newcommand{\bcenter}{\begin{center}}
\newcommand{\ecenter}{\end{center}}

\def\IB{\relax\hbox{$\inbar\kern-.3em{\rm B}$}}
\def\IC{\relax\hbox{$\inbar\kern-.3em{\rm C}$}}
\def\ID{\relax\hbox{$\inbar\kern-.3em{\rm D}$}}
\def\IE{\relax\hbox{$\inbar\kern-.3em{\rm E}$}}
\def\IF{\relax\hbox{$\inbar\kern-.3em{\rm F}$}}
\def\IG{\relax\hbox{$\inbar\kern-.3em{\rm G}$}}
\def\IGa{\relax\hbox{${\rm I}\kern-.18em\Gamma$}}
\def\IH{\relax{\rm I\kern-.18em H}}
\def\IK{\relax{\rm I\kern-.18em K}}
\def\IL{\relax{\rm I\kern-.18em L}}
\def\IP{\relax{\rm I\kern-.18em P}}
\def\IR{\relax{\rm I\kern-.18em R}}
\def\IZ{\relax\ifmmode\mathchoice
{\hbox{\cmss Z\kern-.4em Z}}{\hbox{\cmss Z\kern-.4em Z}}
{\lower.9pt\hbox{\cmsss Z\kern-.4em Z}}
{\lower1.2pt\hbox{\cmsss Z\kern-.4em Z}}\else{\cmss Z\kern-.4em Z}\fi}
\def\II{\relax{\rm I\kern-.18em I}}

\def\sCC{{\kern 0.27em\vrule height1.45ex width0.03em depth0em
          \kern-0.30em\rm C}}
\def\C{{\mathchoice
  {\sCC}
  {\sCC}
  {\kern 0.225em \vrule height1.05ex width0.025em depth0em \kern-0.25em \rm C}
  {\kern 0.180em \vrule height0.78ex width0.02em depth0em \kern-0.2em \rm C}
        }}
\def\sHH{{\rm I\kern-.16em{}H}}
\def\H{{\mathchoice
  {\sHH}
  {\sHH}
  {\rm I\kern-.13em{}H}
  {\rm I\kern-.13em{}H} }}
\def\sNN{{\rm I\kern-.16em{}N}}
\def\N{{\mathchoice
  {\sNN}
  {\sNN}
  {\rm I\kern-.12em{}N}
  {\rm I\kern-.10em{}N} }}
\def\sPP{{\rm I\kern-.16em{}P}}
\def\P{{\mathchoice
  {\sPP}
  {\sPP}
  {\rm I\kern-.12em{}P}
  {\rm I\kern-.10em{}P} }}
\def\sQQ{{\kern 0.27em \vrule height1.45ex width0.03em depth0em
          \kern-0.30em \rm Q}}
\def\Q{{\mathchoice
        {\sQQ}
        {\sQQ}
  {\kern 0.225em \vrule height1.05ex width0.025em depth0em \kern-0.25em \rm Q}
  {\kern 0.180em \vrule height0.78ex width0.020em depth0em \kern-0.20em \rm Q}
        }}
\def\sRR{{\rm I\kern-0.16em{}R}}
\def\R{{\mathchoice
  {\sRR}
  {\sRR}
  {\rm I\kern-0.12em{}R}
  {\rm I\kern-0.10em{}R} }}
\def\sZZ{{\rm Z\kern-0.32em{}Z}}
\def\Z{{\mathchoice
  {\sZZ}
  {\sZZ} 
  {\rm Z\kern-0.3em{}Z}     
  {\rm Z\kern-0.25em{}Z} }}  
\def\ZZZ{{\rm Z\kern-0.24em{}Z}}
\def\sII{{\rm I\kern-0.16em{}I}}
\def\I{{\mathchoice
  {\sII}
  {\sII}
  {\rm I\kern-0.12em{}I}
  {\rm I\kern-0.10em{}I} }}

\def\Tr{{\rm Tr}}

\def\inbar{\,\vrule height1.5ex width.4pt depth0pt}
\font\cmss=cmss10 \font\cmsss=cmss10 at 7pt

\def\smiley{\hbox{\large$\bigcirc$\hspace{-0.80em}\raise.2ex
\hbox{$\cdot\cdot$}\kern-.61em\lower.2ex\hbox{\scriptsize$\smile$}}\ }
\def\frowny{\hbox{\large$\bigcirc$\hspace{-0.80em}\raise.2ex
\hbox{$\cdot\cdot$}\kern-.635em\lower.2ex\hbox{\scriptsize$\frown$}}\ }

\def\I{{\rlap{1} \hskip 1.6pt \hbox{1}}}

\makeatletter
\let\hangafter\@hangfrom
\makeatother

\newcommand{\drawsquare}[2]{\hbox{%
\rule{#2pt}{#1pt}\hskip-#2pt
\rule{#1pt}{#2pt}\hskip-#1pt
\rule[#1pt]{#1pt}{#2pt}}\rule[#1pt]{#2pt}{#2pt}\hskip-#2pt
\rule{#2pt}{#1pt}}

\newcommand{\fund}{\raisebox{-.5pt}{\drawsquare{6.5}{0.4}}}
\newcommand{\Ysymm}{\raisebox{-.5pt}{\drawsquare{6.5}{0.4}}\hskip-0.4pt%
        \raisebox{-.5pt}{\drawsquare{6.5}{0.4}}}
\newcommand{\Yasymm}{\raisebox{-3.5pt}{\drawsquare{6.5}{0.4}}\hskip-6.9pt%
        \raisebox{3pt}{\drawsquare{6.5}{0.4}}}
\newcommand{\antifund}{\overline{\fund}}

\newcommand{\bYsymm}{\overline{\Ysymm}}

\newcommand{\be}{\begin{equation}}
\newcommand{\ee}{\end{equation}}
\newcommand{\bea}{\begin{eqnarray}}
\newcommand{\eea}{\end{eqnarray}}
\newcommand{\bean}{\begin{eqnarray*}}
\newcommand{\eean}{\end{eqnarray*}}

\newcommand{\beqa}{\begin{eqnarray}}
\newcommand{\eeqa}{\end{eqnarray}}
\newcommand{\id}{\bf 1}

\def\Tr{{\rm Tr \,}}

\def\CM {{\cal M}}
\def\FH {{\bf H}}

\preprint{CERN-PH-TH/2006-146 \\ PUPT-2206 \\ IFT-UAM/CSIC-06-37
  \\ hep-th/0607218}

\title{Non-supersymmetric Meta-stable Vacua from Brane 
Configurations}

\author{Sebasti\'an Franco${}^1$, I\~naki Garcia-Etxebarria${}^2$  and 
Angel M. Uranga ${}^{3,2}$

\\
~\\
${}^1$Joseph Henry Laboratories, Princeton University,
Princeton, NJ  08544, USA 
\footnote{Research supported by the United States Department of Energy, under contract DE-FG02-91ER-40671.}
\\ 
${}^2$
Instituto de F\'{\i}sica Te\'orica, C-XVI, UAM, 28049 Madrid, Spain
\footnote{Research supported by the Gobierno Vasco PhD fellowship 
program and the Marie Curie EST program.}
\\
${}^3$ PH-TH Division, CERN, CH-1211 Geneva 23, Switzerland \\
\footnote{Research supported by CICYT, Spain, under project 
FPA-2003-02877, and the RTN networks MRTN-CT-2004-503369 `The Quest for 
Unification: Theory confronts Experiment' and MRTN-CT-2004-005104 
`Constituents, Fundamental Forces and Symmetries of the Universe'.}
\email{sfranco@feynman.princeton.edu, 
Inaki.Garcia.Echebarria@cern.ch, angel.uranga@cern.ch} \\
}

\abstract{
We construct configurations of NS-, D4-, and D6-branes in type IIA string 
theory, realizing the recently discussed non-supersymmetric meta-stable 
minimum of 4d $\mathcal{N}=1$ $SU(N_c)$ super-Yang-Mills theories with massive 
flavors. We discuss their lift to M-theory and the mechanism of 
pseudo-moduli stabilization. We extend the construction to many other 
examples of meta-stable minima, including the $SO/Sp$ theories, 
$SU(N_c)$ with matter in two-index tensor representations, and to a chiral 
gauge theory.
}
\begin{document}

\section{Introduction}

The realization in \cite{Intriligator:2006dd} that simple theories like 
4d $\mathcal{N}=1$ SYM theories with massive flavors contain non-supersymmetric 
local meta-stable minima has triggered a lot of interest in this 
phenomenon, and in particular in its realization in string theory
\footnote{Meta-stable non-supersymmetric vacua had been 
discussed in the supersymmetry model building literature, see e.g. 
\cite{Dimopoulos:1997ww}.}. One 
possible embedding is by considering configurations of D-branes at a 
Calabi-Yau singularity \cite{Franco:2006es}
(for earlier related work on supersymmetry breaking in this kind of 
configurations, see 
\cite{Berenstein:2005xa,Franco:2005zu,Bertolini:2005di}), or wrapped on 
suitable cycles at a local Calabi-Yau \cite{Ooguri:2006pj} \footnote{Gauge sectors of the kind described in \cite{Intriligator:2006dd} 
have been embedded also in heterotic compactifications in \cite{Braun:2006em,Braun:2006da}. 
However, the existence of local meta-stable minima in these constructions, 
where gravity is not decoupled, remains an open question.}. This has led 
to interesting steps in the construction of string theory models of gauge 
mediated supersymmetry breaking \cite{Garcia-Etxebarria:2006rw} (along the 
lines in \cite{Diaconescu:2005pc}).

In this paper we explore a different realization, in terms of type IIA 
configurations of NS-branes with D4-branes suspended between them, in the 
presence of D6-branes, and of their lifts to M-theory. The realization of 
$\mathcal{N}=1$ SYM theories with flavors in this setup, and the realization of
Seiberg duality, are standard \cite{Elitzur:1997fh} (see 
\cite{Giveon:1998sr} for a review), so they provide a solid starting point 
for the discussion. We use these tools to construct the non-supersymmetric 
local meta-stable minimum of SYM theories with unitary, orthogonal or 
symplectic gauge groups (with massive flavors).

The lift to M-theory of the configurations corresponding to the 
supersymmetric vacua of these theories has also been extensively discussed 
(see e.g. \cite{Hori:1997ab,Witten:1997ep,Brandhuber:1997iy}
for $SU(N_c)$ SYM). In these cases, the configuration lifts to a single 
smooth M5-brane wrapped on a holomorphic curve, which encodes important 
information concerning the non-perturbative infrared dynamics of the theory. 

A natural question is whether the information about the low energy 
dynamics of the local meta-stable minima is also encoded in the M5-brane curve.
We describe the main properties of the M-theory lift of the type IIA 
configuration realizing this minimum.
The lift corresponds to a reducible M5-brane geometry, with two 
components which are holomorphic in different complex structures of the 
underlying geometry (which is a Taub-NUT hyper-K\"ahler geometry), and hence 
are volume-minimizing by themselves. One of the components has a number of 
free parameters, which correspond to pseudo-moduli of the effective field 
theory. The mechanism that 
lifts the pseudo-moduli is not encoded in the geometry of the curve, but rather 
these flat directions are removed only when the interaction 
between the two components (described in the large distance regime in 
terms of exchange of gravitons and 3-form fields) is taken into account.
Hence, the 1-loop stabilization in the field theory maps to a process 
beyond the M5-brane probe approximation in the M-theory configuration.
This problem is very difficult with present techniques, hence we simply 
sketch the main points, hoping for further progress in the future.

The realization of known non-supersymmetric local meta-stable minima in 
terms of brane configuration leads to a precise identification of the key 
ingredients in this phenomenon. This allows for many generalizations, and 
we present explicit construction illustrating just a few. We provide the 
type IIA brane configurations corresponding to new non-supersymmetric 
local meta-stable minima in $SU(N_c)$ theory with non-chiral matter in 
symmetric or antisymmetric representations (plus massive flavors), and in a chiral $SU(N_c)$ theory with chiral multiplets in the antisymmetric, 
conjugate symmetric and fundamental representations (plus massive 
flavors).

The paper is organized as follows. In Section \ref{iss} we review the 
field theory description of the non-supersymmetric meta-stable vacua
of \cite{Intriligator:2006dd}. In Section \ref{typeiia} we describe the
type IIA configuration realizing the non-supersymmetric vacuum in the 
$SU(N_c)$ theory with massive flavors, and discuss its classical 
properties. In Section \ref{mlift} we describe the M-theory lift of this 
configuration and discuss the physics encoded (and not encoded) in the 
M5-brane curve. In Section \ref{section_pm_stabilization} we describe the 
physical mechanism lifting the classical pseudo-moduli of the non-supersymmetric 
vacuum, and its realization in string/M -theory. In Section \ref{orient} we 
introduce O4-planes in the type 
IIA configurations to realize the  non-supersymmetric vacua in the $SO$ 
and $Sp$ theories. Further generalizations are constructed 
in Section \ref{general}. Section \ref{conclu} contains our final remarks.

As we were finishing this paper \cite{Ooguri:2006bg} appeared, which
overlaps with the results of Sections \ref{typeiia} and
\ref{section_pm_stabilization}. After we finished this paper
\cite{Bena:2006rg} appeared, which clarifies some aspects of the
discussion in Section \ref{mlift}.

\section{The ISS model}
\label{iss}
                                                                                
In this Section we sketch the analysis in \cite{Intriligator:2006dd}
to determine the existence of meta-stable vacua in $\mathcal{N}=1$ $SU(N_c)$ SYM
with massive flavors. We refer the reader to this reference for details.

Consider $SU(N_c)$ SYM with $N_f$ massive flavors $Q$, ${\tilde Q}$ with 
mass much smaller than $\Lambda$, the dynamical scale of the gauge theory. 
Since the analysis is carried out in the dual theory, we work on the free 
magnetic range $N_c+1 \leq N_f < {3\over 2} N_c$ so that the latter is IR free 
and the K\"ahler potential is under control in the small field region.

The dual theory is $SU(N)$ SYM with $N=N_f-N_c$, with $N_f$ flavors $q$,
${\tilde q}$ and the mesons $\Phi$. They transform as
$(\fund,\antifund,1)$, $(\antifund, 1, \fund)$, $(1,\fund,\antifund)$
under the $SU(N)\times SU(N_f)\times SU(N_f)$ color and flavor symmetry.
                                                                                
The superpotential is of the form
\beqa
W\, =\, h \, \Tr\, (\,q\, \Phi\, {\tilde q}\,)\, -\, h\mu^2 \Tr \, \Phi
\eeqa
(where the traces run over flavor indices).
                                                                                
This theory breaks supersymmetry at tree level due to the F-term of $\Phi$
(the so-called rank condition). There is classical moduli space of minima,
parametrized by the vevs
\beqa
\Phi\, =\, \pmatrix{ 0 & 0\cr 0 & \Phi_0}
\qquad q=\pmatrix{\varphi_0\cr
0},\qquad \tilde q^T=\pmatrix{\tilde \varphi_0\cr 0},\qquad
{\rm with}\,\,\,\,
\tilde\varphi_0\varphi_0 = \mu^2\id_{N}.
\label{classpure}
\eeqa
The computation of the Coleman-Weinberg one-loop effective potential 
shows that all pseudo-moduli (classical flat directions not corresponding 
to Goldstone directions) are lifted in the one-loop effective potential, 
and that the maximally symmetric point in the classical moduli space
\beqa
\Phi_0=0,\qquad \varphi_0=\tilde \varphi_0=\mu\id_{N},
\eeqa
is a minimum of the one-loop effective potential. 

As mentioned in \cite{Intriligator:2006dd}, in the case of different 
flavor masses the local minimum is obtained by setting the $N$ non-zero 
dual quark vevs equal to the $N$ largest masses. If a dual quark vev is 
set to be one of the $N_c$ smallest masses, the configuration is unstable 
already at the classical level, due to the appearance of a negative mass 
squared mode (which triggers the rolling to the correct minimum). 
                                                       
\medskip
                         
The $SU(N)$ gauge dynamics is IR free and hence not relevant in the small
field region, but it is crucial in the large field region. In fact, it
leads to the appearance of the $N_f-N$ supersymmetric vacua predicted by
the Witten index in the electric theory. In the large field
region of $\Phi$ vevs, $|\mu|\ll |\langle h\Phi \rangle|$, the $N_f$
flavors are very massive, and we recover pure $SU(N)$ SYM dynamics, with 
a
dynamical scale $\Lambda'$ given by
\beqa
\Lambda'^{3N} \, =\, \frac{h^{N_f} \, \det \Phi}{\Lambda^{N_f-3N}}
\eeqa
where $\Lambda$ is the Landau pole scale of the IR free theory. The
complete superpotential, including the non-perturbative $SU(N)$ 
contribution
is
\beqa
W = N\, (\, h^{N_f}\, \Lambda^{-(N_f-3N)}\, \det\,\Phi\, )^{1/{N}}\, -\, h
 \mu^2\, \Tr\,\Phi
\eeqa
This superpotential leads to $N_f-N$ supersymmetric minima at
\beqa
\langle h \Phi\rangle\, =\, \Lambda _m\ \, \epsilon ^{2N/(N_f-N)}\, 
{\id_{N_f}} \, =\, \mu {1\over \epsilon ^{(N_f-3N)/(N_f-N)}}{\id_{N_f}},
\eeqa
where $\epsilon \equiv {\mu \over \Lambda}$. In the regime $\epsilon\ll
1$,
the vevs are much smaller than the Landau pole scale, and the analysis can
be trusted. Notice also that these minima sit at $\left|
\langle h\Phi \rangle \right|
\gg \left| \mu \right|$, hence very far from the local non-supersymmetric
minimum. This separation and the height of the potential barrier
make the meta-stable SUSY breaking vacuum long-lived.

\section{Type IIA configuration}
\label{typeiia}

An efficient way to realize supersymmetric gauge field theories in string 
theory is to embed them as the effective gauge theory on the world-volume 
of configurations of D- and NS-branes. These Hanany-Witten setups 
\cite{Hanany:1996ie} have been successfully employed in the study of 
four-dimensional gauge theories with $\mathcal{N}=2$ and $\mathcal{N}=1$ supersymmetry (see
e.g. \cite{Elitzur:1997fh,Witten:1997sc,Hori:1997ab,Witten:1997ep,
Brandhuber:1997iy} and \cite{Giveon:1998sr} 
for a review with more complete references). In this Section we describe 
the type IIA brane configuration that corresponds to the ISS 
non-supersymmetric 
local minimum, and understand some of its classical properties.

\begin{figure}[ht]
  \epsfxsize = 14cm
  \centerline{\epsfbox{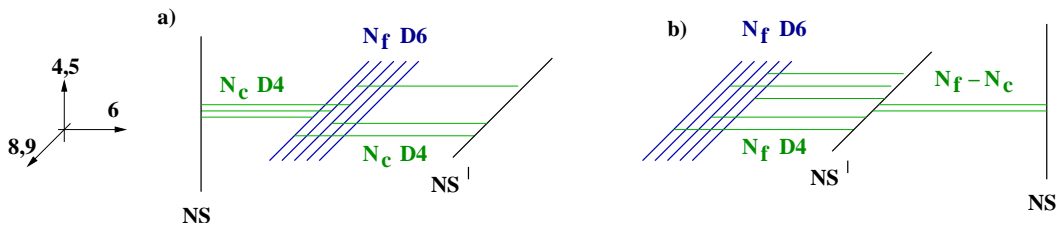}}
  \caption{The type IIA brane configurations for $SU(N_c)$ SYM with $N_f$ 
(massless) flavors (a) and its Seiberg dual theory (b).}
  \label{sym1}
\end{figure}

A convenient starting point is the type IIA brane configuration of $\mathcal{N}=1$ 
$SU(N_c)$ SYM with $N_f$ flavors
\footnote{This configuration can be obtained from the one describing 
$\mathcal{N}=2$ SQCD with $SU(N_c)$ gauge group by rotating the NS' 
(originally parallel to the NS) \cite{Barbon:1997zu}. The relative
angle $\theta$ between the NS and NS' branes dictates the mass of the 
adjoint chiral superfield $|\mu(\theta)|=\tan \theta$. In the limit
in which the NS and NS' become orthogonal $\mu \rightarrow \infty$.
Integrating out the adjoint chiral superfield we are left with 
$\mathcal{N}=1$ $SU(N_c)$ SQCD with vanishing superpotential. Although 
this viewpoint is useful in the derivation of the M-theory lift of the 
type IIA 
configuration, in this section we directly discuss the final rotated 
configuration.}. We consider one NS-brane stretching 
along the coordinates 012345, one NS-brane (denoted NS'-brane) stretching 
along 012389, $N_c$ color D4-branes stretching along 0123 and suspended in 
$x^6$ between the NS- and NS'-branes, and $N_f$ flavor D6-branes 
stretching along 0123789. See \cite{Elitzur:1997fh,Giveon:1998sr} for more 
details. We consider the configuration for zero flavor masses (namely, 
the D6's have all the same position as the NS' in 45). The configuration 
is shown in Figure \ref{sym1}a.

Notice that the D4-branes can in general split in pieces as they are able 
to end on the D6-branes. Notice also the familiar s-rule 
\cite{Hanany:1996ie} which forbids that there are at most one
D4-brane piece connecting the NS-brane with a given D6-brane. The number 
of D4-brane pieces connecting the NS'-brane with a given D6-brane is on 
the other hand arbitrary.

Following the operations in \cite{Elitzur:1997fh,Giveon:1998sr}, it is 
straightforward to obtain the brane configuration describing the Seiberg 
dual theory. Sketchily, one considers moving the NS across the D6-branes 
(process in which the $N_c$ finite D4-brane pieces joining them disappear, 
and $N_f-N_c$ new finite D4-brane pieces appear), and then across the NS'. 
The final configuration is shown in Figure \ref{sym1}b. Notice the 
familiar realization of the meson vevs as the position in $8,9$ of the 
$N_f$ D4-branes pieces suspended between the D6-branes and the NS'-brane.

\subsection{The SUSY breaking minimum}

Let us now consider the type IIA configurations and the above processes in 
the 
presence of non-zero flavor massless, by moving off the D6-branes in the 
directions 45. Consider for simplicity the case where all flavor masses 
are equal. 

The introduction of flavor masses corresponds in the magnetic field
theory to the introduction of the linear term in the mesons that triggers 
supersymmetry breaking. Recall that there is a non-supersymmetric set of 
vacua, where the dual quarks have non-trivial vevs (fixed by the flavor 
masses), and which is parametrized by pseudo-moduli encoded in an 
$N_c\times N_c$ block of the mesonic matrix. 

These features are nicely reproduced by 
the type IIA configuration. When the D6-branes are moved off in 45, 
the $N_f-N_c$ D4-branes joining them to the NS-brane move along 45 and 
maintain the same supersymmetry. However, the $N_f$ D4-brane pieces 
joining them to the NS'-brane misalign with respect to them, leading to a 
non-trivial F-term. The F-term can be partially canceled by recombining 
$N_f-N_c$ of such D4-branes with the D4-branes joining the D6- and the 
NS-branes. This recombination corresponds to the fact that $N=N_f-N_c$ 
entries in the dual quarks acquire non-zero vevs to minimize the F-term.  
Notice that the appropriate breaking of the global symmetry is nicely 
reproduced. For shortness, we sometimes denote D4'-branes the 
D4-branes suspended between the D6- and the NS'-brane. 

\begin{figure}[ht]
  \epsfxsize = 7cm
  \centerline{\epsfbox{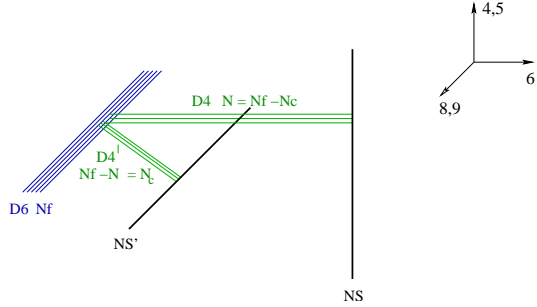}}
  \caption{Type IIA brane configuration corresponding to the SUSY breaking minimum.}
  \label{SUSY_breaking_HW}
\end{figure}

The configuration for the supersymmetry-breaking configuration is shown in 
\fref{SUSY_breaking_HW}
\footnote{This 
configuration is the starting point of our studies. It has been well 
investigated in the past in order to study the deformation 
corresponding to adding terms linear in the mesons to the magnetic 
superpotential (see for example \cite{Giveon:1998sr}. The only new ingredient is to allow the  rank of 
the quark mass matrix of the electric theory (linear couplings in the magnetic 
dual) to be larger than $N_c$. This brane setup was discussed by various 
attendants to a group meeting at the Institute for Advanced Study.}.
Notice that the $N_f-N_c$ D4'-branes joining the D6-branes to the 
NS'-brane are free to move in the directions $8,9$, hence reproducing (most 
of) the classical moduli space of non-supersymmetric vacua of the field 
theory (notice that the pseudo-modulus $\theta$ is not manifest in the 
geometry \footnote{The pseudo-modulus $\theta$ is defined such that
 $\langle \varphi_0 \rangle=\mu e^{\theta}\id_{N}$ and
$\langle \tilde{\varphi}_0 \rangle=\mu e^{-\theta}\id_{N}$ \cite{Intriligator:2006dd}.}).

\medskip

It is very interesting to consider more general situations, with arbitrary 
non-zero masses. Recall that in the field theory analysis in 
\cite{Intriligator:2006dd}, the non-supersymmetric vacua are obtained when 
the vevs for the dual quarks $\varphi_0$, ${\tilde \varphi}_0$ 
are given by the $N$ largest masses (out of 
the $N_f$ mass parameters). In configurations where some vev is given by 
one of the $N_c$ smallest masses, a classically unstable mode appears.

\begin{figure}[ht]
  \epsfxsize = 7cm
  \centerline{\epsfbox{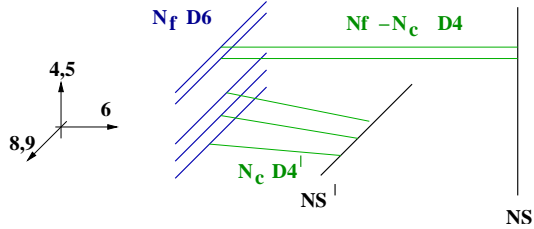}}
  \caption{The non-supersymmetric type IIA configuration reproducing the 
non-supersymmetric ISS field theory minimum for arbitrary flavor masses.}
  \label{differentmass}
\end{figure}

This behavior is easily reproduced by the type IIA configuration. The 
different flavor masses correspond to different $4,5$ positions for the 
different D6-branes. The brane setup corresponding to the classical 
non-supersymmetric configuration suggested in ISS is shown in Figure 
\ref{differentmass}.
In this configuration, the $N_c$ D4-branes connected to the NS'-brane end 
on the $N_c$ D6-branes which are closest (i.e. those associated with the 
$N_c$ smallest mass parameters). This is in order to minimize the energy 
of the configuration. In addition, the remaining $N$ D4-branes 
connected to the NS-brane, end on the farthest $N$ D6-branes (i.e. those 
associated with the $N$ largest mass parameters). The reason for this is 
clarified in the next paragraph. Recall that the position 
in $4,5$ of these D4-branes is related to the dual quark vevs, so we have 
indeed found the configuration realizing the ISS vacuum for different 
masses.

It is now easy to realize what goes wrong if one considers the 
configuration where a dual quark vev is given by one of the $N_c$ smallest 
masses. Clearly, there is one D6-brane on which two D4-brane pieces (one 
connecting to the NS- and other to the NS'-branes) coincide. Since these 
D4-branes are non-supersymmetric with respect to each other, an open 
string tachyon develops at their intersection. This is precisely the 
unstable mode which appears in the field theory analysis . Notice 
that the stretched open string leading to the tachyon is a component of 
the meson field $\Phi$, in agreement with this interpretation \footnote{
One may be surprised by the fact that an open string tachyon is captures 
by a field theory analysis. In fact, similar phenomena occur in other 
non-supersymmetric tachyonic D-brane configurations, in the regime of 
small supersymmetry breaking, see e.g. \cite{Gava:1997jt}.}.
Notice also that this tachyon only appears at the origin in the mesonic 
(pseudo)moduli space, in agreement with the field theory analysis. However 
one cannot try to avoid this classical instability by moving off in 
the (pseudo)moduli space, since quantum corrections (see later) lift it
dynamically pushing the configuration towards the origin.

\medskip

The above brane configuration can also be used to study the situation of 
SQCD with $N_{f,0}$ massless and $N_{f,1}$ massive flavors. These models 
have been discussed in \cite{Franco:2006es} from the field theory 
viewpoint. In particular, it was shown that the magnetic dual exhibits 
supersymmetry breaking by the rank condition for $N_{f,0}<N_c$, and that 
in this case one does not have a meta-stable minimum, but rather a saddle 
point with a runaway direction parametrized by the mesons formed by the 
massless quarks (and which becomes the runaway triggered by the 
Affleck-Dine-Seiberg superpotential in the large field region). 

The above brane configuration provides a simple explanation of these 
facts. We consider the type IIA brane configuration in which $N_{f,0}$ 
D6-branes sit at the origin in the directions $4,5$. If $N_{f,0}<N_c$, 
then $N=N_f-N_c<N_{f,1}$ and there are some of the $N_{f,1}$ D6-branes 
associated with non-zero masses which are not endpoints of the $N$ D4-branes 
in the D4/NS5 
system. These D6-branes can be used as endpoints of the D4-branes in the 
D4/NS' system and lead to non-supersymmetric configurations. If on the 
other hand $N_{f,0}>N_c$, then $N>N_{f,1}$ and the $N$ D4-branes in the 
D4/NS system occupy all the $N_{f,1}$ massive flavor D6-branes (and some 
more). 
Hence the D4-branes in the D4/NS' system are forced to end on the massless 
flavor D6-branes, leading to a final supersymmetric configuration. Thus 
one reproduces the above mentioned condition to have rank supersymmetry 
breaking.

\medskip

The above discussion leads to an important observation. There is a 
dynamical `s-rule' in the non-supersymmetric configurations of our 
interest, which prevents a D4-brane and a D4'-brane to end on the same 
D6-brane. Although more manifest in the case of different masses, this 
conclusion is general and valid in the case of equal masses. This has an 
important implication on the structure of the M5-brane describing the 
M-theory lift of our type IIA configurations.

\subsection{The SUSY breaking minimum in the electric theory}

\begin{figure}[ht]
  \epsfxsize = 12cm
  \centerline{\epsfbox{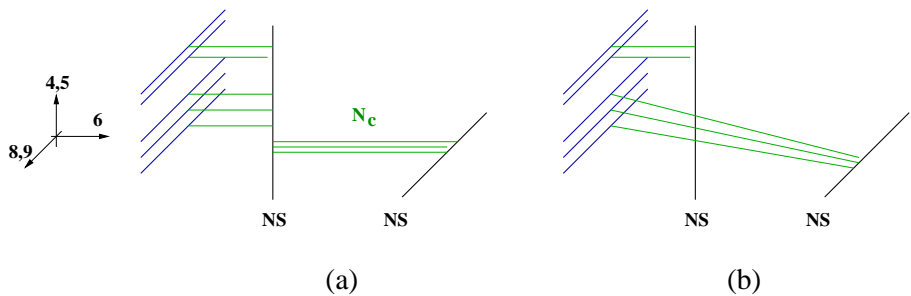}}
  \caption{Figure (a) shows the brane configuration 
describing the supersymmetric minimum of the electric theory. Figure (b) 
shows the supersymmetry breaking meta-stable vacua in the 
brane configuration realizing the electric theory.}
  \label{minimum2}
\end{figure}

Once we have identified the structure of the supersymmetry breaking 
meta-stable minimum in the brane configuration realizing the magnetic 
theory, it is possible to obtain it in the brane realization of the 
electric theory. This is simply obtained by undoing the Seiberg duality, 
namely by crossing back the NS'- and NS-branes. The resulting 
configuration is shown in Figure \ref{minimum2}.b. The supersymmetric 
configuration corresponding to the supersymmetric minima of the theory is 
shown in Figure \ref{minimum2}.a. 

\subsection{Longevity of the meta-stable SUSY breaking vacuum}

As discussed in \cite{Intriligator:2006dd},  the longevity of the 
meta-stable SUSY breaking vacuum depends on its distance to the SUSY vacua 
in field space and the height of the potential barrier separating them. 
Both of them can be estimated by considering a simple trajectory 
connecting the minima.

The separation between vacua is determined by the expectation value of
$\Phi$ at the supersymmetric minimum, the type IIA brane setup
provides a simple visualization of the barrier height. The
$\varphi_0=\tilde{\varphi}_0=0$ point corresponds to no recombination
of the D4-branes. The increase in length of the branes accounts for
the additional potential energy.

\begin{figure}[ht]
  \epsfxsize = 7cm
  \centerline{\epsfbox{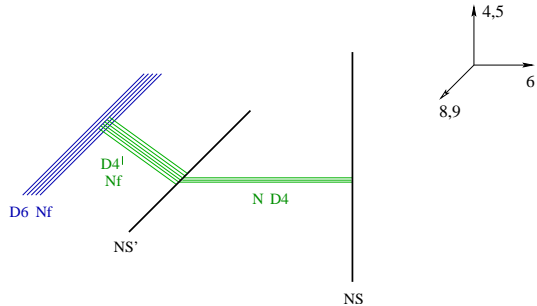}}
  \caption{Brane configuration describing the 
$\varphi_0=\tilde{\varphi}_0=0$ point used to estimate the height of the potential barrier.}
  \label{barrier_HW}
\end{figure}

\section{M-theory lift}
\label{mlift}

In the realization of 4d gauge theories using type IIA brane 
configurations, many interesting properties of the field theory are 
unveiled by lifting the configurations to M-theory 
\cite{Witten:1997sc,Hori:1997ab,Witten:1997ep,Brandhuber:1997iy} 
(see \cite{Giveon:1998sr} for a review). 
The supersymmetric vacua of the theory are easily determined in M-theory. Theories that exhibit dynamical supersymmetry breaking
have been studied in this context in \cite{deBoer:1998by}.
In this section we describe the 
M-theory lift of our type IIA configurations. For simplicity we focus on 
the situation where all flavor masses are equal 
(generalization to different masses is straightforward when 
the $N$ largest masses are arbitrary, but the $N_c$ D6-branes corresponding to the smallest masses are coincident, see later).

\subsection{The factorization}

In the lift to M-theory, D4-branes and the NS- or NS'-branes on which they 
end become different parts of a single smooth M5-brane, wrapped on a 
complex curve in the ambient space (which is given by an $N_f$-centered 
Taub-NUT geometry, corresponding to the M-theory lift of the $N_f$ 
D6-branes). In this section we investigate what is the M-theory configuration that describes the SUSY breaking meta-stable minimum.
From the structure of the type IIA configuration corresponding to the non-supersymmetric vacuum, one can draw 
an important conclusion: the M5-brane curve in the lift of the 
configuration is split into two components. This follows from the fact 
that the D4-branes ending on the NS-brane are completely 
disconnected from the D4-branes ending on the NS'-brane. This implies that 
the D4/NS system and the D4'/NS' lift to two independent M5-brane curves.

Notice moreover that in the case of equal masses each system by itself 
preserves some supersymmetry (more precisely, in order for this to happen the D6-branes connected to D4'-branes must be coincident). Namely, the D4/NS system preserves 8 
supercharges in the presence of the D6-branes, while the D4'/NS' system 
preserves 4 supercharges in the presence of the D6-branes \footnote{For 
general masses, the former system is still supersymmetric, while the 
latter breaks all supersymmetries unless the D6-branes for the $N_c$ smallest masses are coincident.}. Notice however that the supersymmetries preserved 
by both systems are not compatible, and the complete configuration breaks 
all supersymmetries.

This structure has a beautiful counterpart in the M-theory lift. The 
geometry in the M-theory lift is given by a Taub-NUT geometry, which is 
hyper-K\"ahler. Therefore it admits a $\IP_1$ of complex structures. The 
two different M5-brane components corresponding to the lifts of the D4/NS 
and the D4'/NS' systems correspond to M5-branes wrapped on two curves 
which are holomorphic in two different complex structures in this 
geometry. The rotation between the two complex structures in which the two 
curves are holomorphic is related to the amount of supersymmetry breaking 
(namely, to the angle between the D4- and D4'-branes in the type IIA 
configuration). Being holomorphic in some complex structure, each 
component is volume minimizing by itself. However, the complete system can 
be regarded as an M5-brane on a singular (i.e. reducible) non-holomorphic 
curve, which is therefore not volume-minimizing as a whole. 

The M-theory state we build reduces at vanishing string coupling to
the type IIA configuration we have studied. Interestingly, it captures
various features of the IIA configuration such as its pseudo-moduli.
Nevertheless, it is important to notice that this M-theory lift
possesses non-holomorphic boundary conditions. The asymptotic behavior
of our M-theory curve differs from the one of the M-theory lift of the
supersymmetric configuration (in particular it does not preserve
supersymmetry even asymptotically). As a result, it cannot be
interpreted as a state with spontaneously broken supersymmetry in a
supersymmetric 4d theory. Instead, it corresponds to a state in a
theory with a Lagrangian that breaks supersymmetry explicitly. This
issue was investigated in detail in \cite{Bena:2006rg}.

\subsection{The curves}

Following the argument above, we are led to describe the M-theory lift in terms of two curves 
which are holomorphic in two different complex structures of the Taub-NUT 
geometry (again, notice that this assumes a supersymmetric D4'/NS' system, 
hence that the corresponding $N_c$ D6-branes are coincident).

In order to consider the lift of the D4/NS system, let us introduce an 
adapted complex structure, in which the corresponding M5-brane curve is 
holomorphic. In fact the system is locally $\mathcal{N}=2$ supersymmetric, hence we 
may stick to the usual conventions for lifts of configurations of 4d $\mathcal{N}=2$ 
theories \cite{Witten:1997sc}. Let us introduce $v=x_4+ix_5$, 
$w=x_8+ix_9$,
and describe the ambient M-theory Taub-NUT geometry as the complex manifold 
\beqa
yz\, =\, [\, \prod_{i=N_c+1}^N\,  (v-\mu_i)\, ] \, (v-\mu)^{N_c}
\eeqa
where $(\mu_{N_c+1},\ldots, \mu_{N_f})$ correspond to the $N$ largest 
mass parameters and $\mu<\mu_i$ is the common mass parameter for the $N_c$ 
lightest flavors. Notice that the mass parameters encode the positions of 
the D6-branes (or of the Taub-NUT centers in M-theory) in the $4,5$ 
directions. 

In these complex coordinates, the holomorphic curve corresponding to the 
D4/NS system has the structure
\beqa
& z \, - \,\prod_{i=N_c+1}^N\,  (v-\mu_i) = 0 \nonumber \\
& w\, =\, 0
\eeqa
Hence, the M5-brane has spikes towards $z\to 0$ at the positions 
$v=\mu_i$. These spikes become the D4-branes upon reduction to type IIA. 
The interpretation for these spikes is that 
$v\to \mu_i$ corresponds to the cycle $yz=0$ in the ambient Taub-NUT. This 
is reducible, and $z=0$, $y$ arbitrary, describe one component, 
corresponding to a spike ending on the Taub-NUT center (from the right).

\medskip

The lift of the D4'/NS' component is also easily described, in the case of 
a common mass for the lightest flavors, on which we are centering. The 
system is locally $\mathcal{N}=1$ supersymmetric, so it is described by a 
holomorphic curve in adapted complex coordinates. Intuitively we
introduce $v'=x_4'+ix_5'$, where $x_4'$ and $x_5'$ parametrize the 2-plane 
orthogonal to the $x_6'$ direction along which the D4'-branes stretch. 
Similarly we need to introduce new complex parameters $\mu_i'$, $\mu'$ 
which encode the positions of the D6-branes in the $v'$ direction.
The mapping of complex coordinates in different complex structures 
(rotated by the $SU(2)$ isometry of the Taub-NUT geometry, equivalently 
the $SO(3)$ rotation in the space parametrized by $4,5,6$), and of different 
complex parameters specifying the D6-brane positions, is somewhat 
technical, but we provide its description in Appendix \ref{appendix_kahler}.

In the complex structure adapted to the D4'/NS' system, the Taub-NUT 
geometry is described by
\beqa
y'z'\, =\, [\, \prod_{i=N_c+1}^N\,  (v'-\mu_i') \, ] \, (v'-\mu')^{N_c}
\eeqa
In these complex coordinates, the holomorphic curve describing the lift of 
the D4'/NS' system is 
\beqa
& z' \, - \, \prod_{i=1}^{N_c} \,(w'-w_i')\, =\, 0 \nonumber \\
& v'\, =\, 0
\eeqa
Namely the M5-brane has spikes of order $N_c$ towards $x_6'\to -\infty$ 
(i.e. $z'\to 0$) at the position $w'=w_i'$. The parameters $w_i'$ 
are free moduli of the holomorphic curve, and correspond to the 
mesonic field theory pseudo-moduli. Hence they remain as flat directions 
even in the M-theory lift.

It would be interesting to describe the lift of the type IIA configuration 
for completely general mass parameters, in particular when the D4'/NS' 
system is non-supersymmetric by itself. We leave this interesting point 
for future work.

\medskip

Being holomorphic in some complex structure, the above curves are 
automatically area-minimizing, and hence indeed correspond to the
a classical stationary configuration for the M5-brane configuration, in 
the probe approximation. We will describe in Section \ref{section_pm_stabilization} the impact 
of 
possible backreaction effects in the curves, and on the pseudo-moduli 
stabilization.

We conclude with a final remark. Notice that the positions of the 
D6-branes in 6 enter in the determination of the complex parameters 
$\mu'$, $\mu_i'$, and hence appear in the expression of the curve.
It is a familiar fact that in supersymmetric vacua such positions are 
hidden parameters of the brane configuration which are not visible 
in holomorphic quantities of the gauge theory. Hence it is not unexpected 
that they pop up as relevant quantities when dealing with a 
non-supersymmetric vacuum. It would be interesting to gain a better 
understanding of the interplay of gauge theory quantities and these 
parameters.

\subsection{A heuristic argument}
\label{section_heuristic}

We would like to conclude with a suggestive heuristic derivation of the 
factorized structure. In the field theory description, the ISS 
construction of the local minimum can be obtained by starting with 
the electric theory, moving on to the magnetic dual and taking the limit in which the gauge interactions vanish. This procedure can be carried out in the 
M-theory description of the configuration, providing independent 
evidence for the factorized structure.

The infrared physics of the electric theory is described by the M5-brane 
wrapped on the holomorphic curve obtained in \cite{Hori:1997ab}, 
\beq
t=w^{N_c-N_f}(w-w_0)^{N_f}
\label{curve_1}
\eeq
\beq
v \, w = m \, w_0
\label{curve_2}
\eeq
with
\beq
w_0=\left({\Lambda_{SQCD}^{3N_c-N_f} \over m^{N_c-N_f}} \right)^{1/N_c}
\label{w0}
\eeq
where $t$ is related to the usual variable used in $\mathcal{N}=2$ 
configurations by a rescaling by $\mu^{N_c}$, with $\mu$ the adjoint mass 
that goes to infinity in the $\mathcal{N}=1$ limit. The value of $w_0$ 
determines the expectation value of the mesons at the supersymmetric vacua. It is straightforward to rewrite $w_0$ in terms of magnetic
quantities.

The two equations defining the curve can be rewritten to clarify how the 
curve relates to the lifts of the NS and NS'. Rewriting \eref{curve_1} 
as
\beq
t-w^{N_c-N_f}(w-w_0)^{N_f}=0
\label{curve_3}
\eeq
we see $N_c$ D4-branes ending on the NS'-brane (from the left). Combining 
\eref{curve_1} 
and \eref{curve_2} we obtain
\beq
\left({m^{N_f-N_c}\over w_0^{N_c}} v^{N_c} \right) t - (m-v)^{N_f}=0
\label{curve_4}
\eeq
showing $N_c$ (resp. $N_f$) D4-branes ending on the NS-brane (from the 
right resp. left).

As mentioned above, in the field theory one constructs the 
non-supersymmetric vacuum by going to the magnetic dual and turning off
the magnetic gauge interactions. In order to heuristically perform this 
operations in terms of the M5-brane curve, it is useful to rewrite 
quantities in terms of magnetic variables, which are related to the 
electric ones by
\beqa
\Lambda_{SQCD}^{3N_c-N_f} \Lambda^{3(N_f-N_c)-N_f}=\hat{\Lambda}^{N_f}
\quad  ; \quad
h=\Lambda/\hat{\Lambda}
\quad ; \quad
\mu^2=-m \,\hat{\Lambda}
\eeqa
We then have
\beq
m=-{h \mu^2 \over \Lambda} \ \ \ \ ; \ \ \ w_0= \left( 
{\Lambda^{N_c+(3N_c-2N_f)}\over (-\mu^2)^{N_c-N_f} h^{N_c}}\right)^{1/N_c}
\eeq
\beq
m \, w_0=\left((-\mu^2)^{N_f} \Lambda^{3N_c-2N_f}\right)^{1/N_c}
\eeq
The non-supersymmetric vacuum appears in the magnetic theory when we take 
the classical limit, turning off the gauge interactions. 
Hence, we are interested in the limit $\Lambda \rightarrow 0$ with $h$ 
and $\mu$ fixed. In this limit
\beq
\begin{array}{cl}
m \sim \Lambda^{-1} & \rightarrow \infty \\
w_0 \sim \Lambda^{1+(3N_c-2N_f)/N_c} & \rightarrow 0 \\
m \, w_0 \sim \Lambda^{(3N_c-2N_f)/N_c} & \rightarrow 0 
\end{array}
\eeq
where we have explicitly used that we are working on the free magnetic
range $N_c+1 \leq N_f < {3\over 2} N_c$. With this, \eref{curve_2}
becomes
\beq
v \, w \rightarrow 0
\eeq
We thus see that the curve splits into two components. One of them 
(associated with the D4'/NS' system in the type IIA configuration)
corresponds to $v=0$ and is given by the limit of \eref{curve_3}. The 
other one (associated with the D4/NS system) has $w=0$ and is given by the 
limit of \eref{curve_4}. These curves have the expected behavior i.e. 
for the first one $t\sim w^{N_c}$ as $w \rightarrow \infty$ and for the 
second one $t \sim v^{N_f-N_c}=v^N$ as $v \rightarrow \infty$. 

The argument is heuristic since the curve cannot completely agree with the 
complete M5-brane curve that we have determined in previous sections.
This is because the holomorphic curve remains holomorphic in the limit. 
However, the naive translation of the ISS field theory construction to the 
M5-brane curve does have a suggestive structure. Indeed the curve 
reproduces the correct structure to the best extent that one can expect from a 
holomorphic curve! Namely, it factorizes into two components which 
have the correct number of spikes ending on the correct NS/NS' fivebranes.
The only caveat is that the two components are holomorphic in the same 
complex structure. Our interpretation is that, since the holomorphic 
curve of the supersymmetric vacuum is insensitive to the positions of the 
Taub-NUT centers in the $x_6$ direction, it reproduces the correct structure for the 
non-susy vacuum in the limit where the centers are sent off to $x_6\to 
-\infty $ (in which it becomes holomorphic). In a sense, it is the only 
regime where the holomorphic curve can be expected to match the M5-brane 
curve of the non-supersymmetric vacuum.

In order to have a closer look at the factorized holomorphic curve arising 
in the limit, it is useful to introduce the rescaled variables 
\beq
\begin{array}{rcl}
\tilde{t} & = & t/w_0 ^{N_c} \\
\tilde{v} & = & v/m \\
\tilde{w} & = & w/w_0  
\end{array}
\eeq
The two components that follow from \eref{curve_3} and \eref{curve_4} become
\beq
\tilde{t}-\tilde{w}^{N_c}=0
\label{curve_5}
\eeq
\beq
\tilde{v}^{N_c} \, \tilde{t}- (1-\tilde{v})^{N_f}=0
\label{curve_6}
\eeq

They represent the two components elongating to infinity with different 
asymptotic behavior in $\tilde{v}$. It is easy to realize that they agree 
with the two components of our M5-brane curve in previous sections, in a 
suitable limit (in which in particular the two complex structures become 
the same).

\section{Pseudo-moduli stabilization}

\label{section_pm_stabilization}

We have shown that the field theory pseudo-moduli correspond to
geometric moduli of type IIA configuration. They moreover remain
flat directions in the M-theory configuration, at least in the
case where the D4'/NS' system is supersymmetric, where we could determine
the structure of the curve. Hence the quantum
gauge theory effects encoded in the M-theory curves do not include the
1-loop correction lifting these accidental flat directions.
                                                                                
In fact it is easy to understand what the mechanism responsible for the
stabilization is. The
fields that contribute to the field theory 1-loop Coleman-Weinberg
potential are classically massive fields whose mass depends on the
pseudo-moduli. In the type IIA brane configuration, the pseudo-moduli are
the geometric positions of D4'-branes in $8,9$. Clearly, the classically
massive fields whose mass depends on these positions are D4-D4' and
D4'-D4' open strings. However, the D4'-D4' open strings are not sensitive
to the breaking of supersymmetry and do not contribute, hence only the
D4-D4' states contribute. The Coleman-Weinberg potential then corresponds
to the annulus diagram with boundaries on the D4- and D4'-branes. 
Hence the lifting of pseudo-moduli is an effect that cannot be detected 
from the study of the D4/NS or D4'/NS' systems in isolation, but which 
arises from their interaction.

Due to the complicated geometry (and the presence of the NS- and 
NS'-branes) this
diagram cannot be computed for arbitrary locations of the D4'-branes. In 
the small distance regime, its result should reproduce the field theory
result. Unfortunately the brane configuration does not seem to provide new 
insights in this regime. On the other hand, in the large distance regime, 
the annulus diagram corresponds to the exchange of supergravity modes 
(graviton, dilaton and 5-form exchange) between the D4- and 
D4'-branes. Being non-supersymmetric, it is expected that the 
gravitational exchange overcomes the RR-form repulsion (which is smaller 
due to the misalignment of the D-branes) and lead to a net attraction, 
which pushes the D4'-branes towards the origin in $8,9$. 
This is the string theory view of the lifting of the pseudo-moduli, in 
the large field region of pseudo-moduli space.

The above discussion can be lifted to M-theory. Each M5-brane component is 
area-minimizing by itself, and the component corresponding to the lift of 
the D4'/NS' system has arbitrary moduli. Their lifting can be described 
in terms of the interaction between the two components, which in the long 
distance regime reduces to graviton and 3-form exchange. This implies that 
the lifting of moduli requires describing the configuration beyond the 
brane probe approximation. In principle, a quantitative computation of 
this effect could be achieved by considering the backreaction of the 
M5-brane component associated with the D4/NS system, and solving the area 
minimization equations, in the backreacted background, for the 
M5-brane component associated with the D4'/NS' system. A similar approach 
was used in \cite{Hori:1997ab}, where the above procedure to 
compute supergravity interactions between different M5-brane components 
provided certain correction to the metric on the Higgs branch of $\mathcal{N}=2$ 
gauge theories. The reduced (in fact, absence of) supersymmetry in our 
present problem clearly suggest that the computation is might be considerably more difficult and beyond the scope of this paper. We hope to come back 
to this point in future work.

\section{Symplectic and orthogonal gauge groups}
\label{orient}

It is straightforward to carry out a similar discussion for the 
non-supersymmetric minima in the $SO(N_c)$ and $USp(N_c)$ theories 
with $N_f$ massive flavors. In fact, the type IIA configurations 
realizing these gauge theories, and their Seiberg duality properties,
have been studied in \cite{Evans:1997hk}. We sketch the new features of 
this construction, referring the reader to this reference (see also
\cite{Giveon:1998sr} for a review) for details.

The construction of the electric 
theories combines the same ingredients as for the $SU(N_c)$ theory 
(namely, $N_c$ D4-branes suspended between an NS- and an NS'-brane, in 
the presence of $N_f$ D6-branes), plus an additional O4-plane, stretching 
along the directions 01236 (i.e. parallel to the D4-branes). The O4-plane 
flips its charge as it crosses the NS- and NS'-branes.

The introduction of the O4-plane pairs up the D4-branes in two sets, 
related by the orientifold symmetry, and reduces their gauge symmetry down to 
$SO(N_c)$ or $USp(N_c)$ when the middle piece of the O4-plane has negative 
or positive charge, respectively. Similarly, the D6-branes pair up and 
reproduce the appropriate global symmetries for these gauge theories. 
Finally notice that for odd $N_c$ the $SO(N_c)$ configuration has an 
unpaired D4-brane on top of the middle part of the O4-plane, while for the 
$USp(N_c)$ configuration this is not possible (see 
\cite{Hori:1998iv,Gimon:1998be} for details). Notice that our notation for 
the number of D-branes is dictated by counting on the covering space. In 
this convention, the O4-plane charge is $\pm 4$ D4-brane charge units.

Seiberg duality is obtained by moving the NS across the D6- and the 
NS'-brane. In the process, there is a change in the number of D4-branes 
which determines the final number $N$ of D4-branes joining the D6-branes 
and the NS-brane (which controls the rank of the Seiberg dual gauge group). 
Since there is a contribution of the O4-plane charge to this 
Hanany-Witten effect, one obtains $N=N_f-N_c+ 4$ for $SO$ and $N=N_f-N_c-4$ for $USp$.

\begin{figure}[ht]
  \epsfxsize = 10cm
  \centerline{\epsfbox{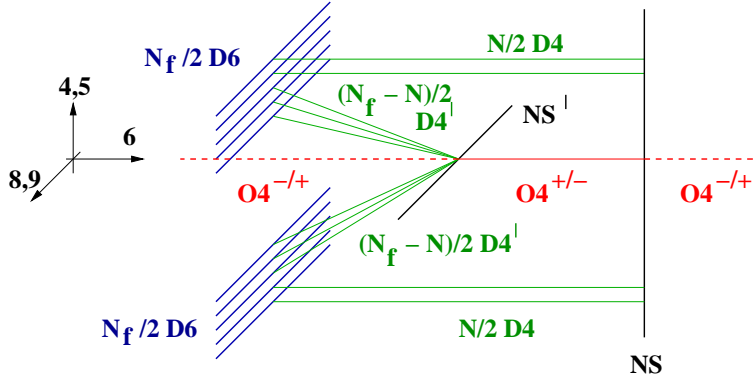}}
  \caption{The type IIA brane configuration for the non-supersymmetric 
minimum of the $SO(N_c)$ and $USp(N_c)$ gauge theories with $N_f$ massive 
flavors. For clarity we have shown the situation for different flavor 
masses. The red line corresponds to the O4-plane, and its change of charge 
as it crosses the NS- and NS'-branes is shown as a change between dashed 
and solid. For the $SO(N_c)$ theory we have $N=N_f-N_c+4$, and it 
corresponds to choosing the O4-plane to have negative charge in the middle 
interval, and positive on the semi-infinite pieces. For the $USp(N_c)$ 
theory we have $N=N_f-N_c-4$ and the O4-plane charge is positive in the 
middle interval.}
  \label{sosp}
\end{figure}

Using these rules one can directly construct the type IIA configuration
corresponding to the non-supersymmetric minima for the $SO$ and $USp$ 
theories described in \cite{Intriligator:2006dd}. It is shown in Figure 
\ref{sosp}. As in the $SU$ case, it is possible to match all classical 
properties of the field theory with geometric properties of this 
configuration.

An important difference is that the D4'/NS' system is non-supersymmetric 
by itself (even in the case of equal flavor masses). In particular one may 
be worried by the fact that we have several D4-branes at angles which 
seemingly intersect (as they reach the NS'-brane) and could potentially 
lead to tachyons. One may think that the presence of the O4-plane imposes 
an orientifold projection that removes them; however, given that the 
tachyonic modes have a matrix structure, they cannot be completely removed 
by such orientifold projection, which at most projects the matrix down to 
the symmetric or antisymmetric components. The key ingredient must 
therefore be the presence of the NS'-brane. Indeed, the presence of such 
object at the coincidence of the D4-branes can prevent the naive claim 
that there is an open string tachyon in the D4-D4 spectrum. In the 
following we assume that this is indeed the case (as suggested by its 
agreement with the field theory picture that no instability exists) and 
proceed to use this additional rule in our subsequent examples. It would 
be interesting to provide further support for it based on computations of 
open string spectra in the near horizon region of NS-branes as in
\cite{Elitzur:2000pq}.

As in the $SU$ case, the construction shows that the lift to M-theory 
is given by an M5-brane wrapped on a reducible curve. An important 
difference is that, since the D4'/NS' system is non-supersymmetric
by itself (even in the case of equal flavor masses), it lifts to an 
M5-brane component which is not holomorphic (in any complex structure).
We leave this discussion for future work.

\section{Generalizations}
\label{general}

The realization of known non-supersymmetric local meta-stable minima in 
terms of brane configurations leads to a precise identification of the key 
ingredients in this phenomenon. In this section we use this ingredients to 
construct non-supersymmetric 
local meta-stable minima in other field theories which admit a 
realization in terms of type IIA brane configurations \footnote{We do not study the supersymmetric vacua, longevity of the meta-stable minimum, etc. This would be an interesting exercise. We consider the similarity of the brane configurations with those in the previous sections makes it clear that these issues will not change much.}. Clearly there are 
many other possibilities, which we leave as an exercise for the reader.

\subsection*{$SU(N_c)$ with non-chiral matter in the $\Ysymm$ or $\Yasymm$}

\begin{figure}[ht]
  \epsfxsize = 10cm
  \centerline{\epsfbox{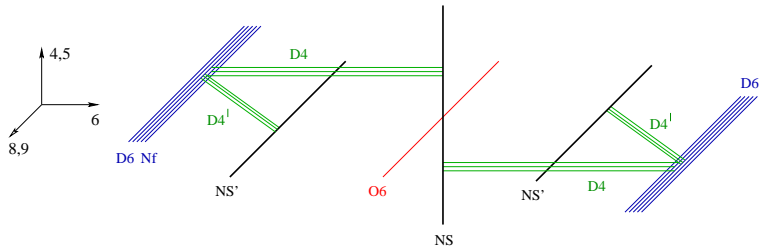}}
  \caption{The type IIA brane configuration describing the 
non-supersymmetric meta-stable minimum in the $SU(N_c)$ theory with 
non-chiral matter in the symmetric or antisymmetric representations 
(corresponding to the choice of positive or negative O6-plane charge), 
plus massive flavors.}
  \label{symtwoindex}
\end{figure}

The type IIA brane realization of the $SU(N_c)$ with non-chiral matter in 
symmetric or antisymmetric (plus additional flavors) has been achieved in
\cite{Landsteiner:1998pb}, by the introduction of O6-planes (along 
0123789). Using the 
ingredients in these configurations, and the basic building blocks of 
non-supersymmetric meta-stable minima, it is straightforward to construct 
a brane configuration realizing a non-supersymmetric meta-stable vacuum in 
these theories. The configuration is shown in Figure \ref{symtwoindex}.

\subsection*{Chiral $SU(N_c)$ theory with chiral multiplets in the 
$\Yasymm+\bYsymm+8\fund$}

We would like to present one example of a chiral theory with a 
non-supersymmetric meta-stable vacuum. Although type IIA brane 
configurations are not particularly well suited for the construction of 
chiral theories (and configurations of D3-branes at singularities may 
provide a better starting point \cite{Franco:2006es}), there is a 
type IIA configuration realizing a chiral $SU(N_c)$ theory with one chiral 
multiplet in the antisymmetric, one in the conjugate 
symmetric, and 8 in the fundamental representations 
\cite{Landsteiner:1998gh,Brunner:1998jr,Elitzur:1998ju}. The 
configuration contains two NS- and one NS'-brane, with an O6-plane 
passing through the latter. The key 
ingredient for producing chirality is that the O6-plane is 
split into two pieces by the NS'-brane, with both pieces carrying different 
charge. In order to cancel an NS'-brane worldvolume tadpole, one needs to 
introduce 8 half D6-branes ending on the latter (see related discussions 
in \cite{Hanany:1997sa,Brodie:1997sz}).

\begin{figure}[ht]
  \epsfxsize = 10cm
  \centerline{\epsfbox{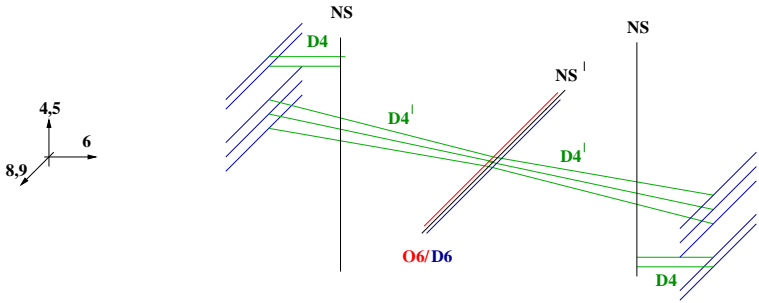}}
  \caption{The type IIA brane configuration describing the 
non-supersymmetric meta-stable minimum in the chiral $SU(N_c)$ theory with 
 matter in antisymmetric, conjugate
symmetric and fundamental representations. Here O6/D6
stands for the system of the split O6-plane and the 8 half D6-branes.}
  \label{chiral}
\end{figure}

The construction of the configuration realizing the non-supersymmetric 
meta-stable vacuum is fairly easy. The only subtlety is that, since the 
NS'-brane must be on top of the O6/D6 system, it is convenient to use the 
configuration where the local minimum  is described in the electric 
theory. The configuration is shown in Figure \ref{chiral}, where O6/D6 
stands for the system of the split O6-plane and the 8 half D6-branes. 

\medskip

As should be clear by now, any type IIA brane configuration can be 
modified to include the basic ingredients involved in the appearance of 
the non-supersymmetric meta-stable minima. It is therefore straightforward 
to generalize to product gauge group theories, etc. We refrain from 
entering a more detailed discussion of these possible generalization, 
leaving them for the interested reader.

A last word of caution is appropriate.
The meta-stable vacua we are studying occur at small expectation values
of fields. In order to be certain of their existence it is thus crucial to have
the K\"ahler potential under control such that we can ensure that it 
remains regular. This is achieved in these examples when the macroscopic 
magnetic description of the theory is IR free. This is easily attainable 
in the $SO(N)$ and $USp(N)$ examples of Section \ref{orient}. From the 
point of view of the electric theory, the free magnetic ranges 
correspond to $N_f<{3\over 2}(N_c-2)$ for $SO(N_c)$ with $N_f$ flavors 
and $N_f<{3\over 2}(N_c+1)$ for $USp(N_c)$ with $2 N_f$ flavors.
We are confident that it is possible to find free magnetic ranges in 
generalizations such as the theories described in this section by 
appropriately tuning the numbers of colors and flavors in the magnetic 
theories. Most of what is currently known about many of these field 
theories comes from their realization by means of D-brane setups, which 
yield no information about the free magnetic and other ranges. On the 
other hand, our experience with the factorization of the M-theory curve in 
Section \ref{section_heuristic} suggests that some understanding of these problems might 
follow from the M-theory lift.

\section{Final remarks}
\label{conclu}

In this paper we have described the construction of type IIA brane 
configurations which realize non-supersymmetric meta-stable vacua of 4d 
$\mathcal{N}=1$ supersymmetric gauge theory. The realization of these vacua for 
the field theory examples in \cite{Intriligator:2006dd} allows us to 
identify the key ingredients of the brane configuration related to the 
existence of these vacua. And hence to generalize the construction to many 
other brane configurations and field theories.

We have also provided a description of the lift of these configurations to 
M-theory, in terms of M5-branes wrapped on a reducible curve. In the 
simplest situation (where some mass parameters are equal) its
components are holomorphic in different complex structures of the 
underlying Taub-NUT geometry. We have argued that a complete understanding 
of the physics of the local minimum, in particular of the lifting of the 
pseudo-moduli, requires a description beyond the brane probe 
approximation, namely taking into account the interaction between the two 
components. The quantitative treatment of this problem thus remains an 
important open issue in these constructions.

It would be interesting to find connections between the constructions we 
have presented in this paper, and the discussion of non-supersymmetric 
meta-stable vacua for gauge theories on systems of D-branes at 
singularities. It is possible that T-duality relations along the lines of
\cite{Uranga:1998vf,Dasgupta:1998su} provide a bridge between both 
languages.

We expect much progress in the understanding of these non-supersymmetric
meta-stable vacua from their realization in string theory.

\section*{Acknowledgments}

We thank Nissan Itzhaki for useful discussions. A.U. thanks M. Gonz\'alez for
encouragement and support.

\appendix

\section{The hyper-K\"ahler structure of Taub-NUT and rotation of the
  holomorphic structure}

\label{appendix_kahler}

\subsection{Hyper-K\"ahler construction of the multicenter Taub-NUT spaces}
Let us describe the (multicenter) Taub-NUT space as a hyper-K\"ahler
quotient, following the reasoning in \cite{Gibbons:1996nt}.
In order to build the
Taub-NUT in this way, we start from the manifold $\CM$ given by $d+1$
copies of $\FH$, where $d$ is the number of centers in our space and
$\FH$ is a copy of ${\bf R}^4$ with flat hyper-K\"ahler metric. Let us
take as coordinates in $\CM$ the quaternions $w$ and $q_a$, where $a$
goes from 1 to $d$.

Now consider the abelian group $G$ of rank $d$ acting on the manifold,
this group is isomorphic to ${\bf R}^d$ locally. The moment map for
this group acting on $\CM$ is given by
\be
\mu_a=\frac{1}{2}{\bf r}_a + {\bf y},
\ee
where ${\bf r}_a=q_a i \bar{q}_a$ (no sum in $a$, and boldface denotes
three dimensional vectors) and
${\bf y} = (w-\bar w)/2$. Under the $a$-th factor of $G$, $q_a$
transforms with a $+1$ $U(1)$ charge, $w$ gets translated and the rest
of the coordinates remain invariant. Let us consider the set of
vectors ${\bf e}_a$. Then we define our Taub-Nut space as:
\be
X=\mu^{-1}({\bf e})/G,
\ee
where the $a$ index is implicit. Namely, we consider all points in
$\CM$ such that their moment maps $\mu_a$ give ${\bf e}_a$, and then
quotient the resulting space by the action of $G$. With the metric
inherited from the flat $\CM$ one gets the multi center Taub-NUT space
with the standard metric:
\be
ds^2 = \frac{1}{4} V d{\bf r}^2 + \frac{1}{4}V^{-1}(d\tau +
{\vec \omega}\cdot d{\bf r})^2,
\ee
with $\nabla\times{\vec \omega} = {\vec \nabla}V$ and
\be
V = 1+\sum_{a=1}^{d}\frac{1}{|{\bf r} - {\bf e}_a|},
\ee
so we can identify the values of the moment maps with the positions of
the centers of the Taub-NUTs.

\subsection{Complex structure for these spaces}
In this section we will try to understand better the structure as a
complex manifold of the space we just built following
\cite{Witten:1997sc}. Recall that a quaternion can be written as
$q=a+ib+jc+kd$, where $i$, $j$ and $k$ satisfy the $SU(2)$ Lie group
algebra, so we can think of them as the Pauli matrices, and $a$, $b$,
$c$ and $d$ are real numbers. This structure reflects the hyper-K\"ahler
nature of the manifold, we can associate choosing a complex structure
(in the $S^2$ of possible complex structures) with privileging $i$,
say, and then decomposing the quaternion q into two complex numbers
$w_1$ and $w_2$ given by:
\bea
w_1 & = & a + i b \\
w_2 & = & c + i d,
\eea
the decomposition being motivated by $q = a+ib+j(c+id)$. We have the
freedom of choosing any combination of $i$, $j$ and $k$ as determining
a complex structure, so we have essentially the freedom of choosing a
direction in ${\bf R}^3$. Note also that the 3-vector structure of the
moment maps also comes from $ijk$, so rotating the directions of the
base space for the Taub-NUT (i.e., ordinary rotations in the type IIA
picture) roughly corresponds to choosing different complex
structures. This will be the basic idea in what follows.

Let us privilege a complex structure and then separate $q_a$ into the
complex variables $y_a$ and $z_a$, and $w$ into $v$ and $v'$, such
that the action of $G$ in these complex variables is given by:
\bea
y_a & \rightarrow & e^{i\theta_a} y_a \\
z_a & \rightarrow & e^{-i\theta_a} z_a \\
v & \rightarrow & v \\
v' & \rightarrow & v' - \sum_{a=1}^d \theta_a.
\eea

Also, when we pick a complex structure, namely a direction in 3 space,
the moment map can be divided into the longitudinal part (a real part
$\mu_{\bf R}$) and a transverse (complex) part $\mu_{\bf C}$. In the
type IIA picture the latter corresponds to the projection of the D6
brane position into the $4,5$ plane in which the NS brane is sitting, and
the former to the $x^6$ position of the brane, which as we will see
below does not appear in the defining equations for the NS factor of
the M theory curve in the complex structure in which it is holomorphic.

The components of $\mu_{\bf C},a$ give the equations:
\be
y_a z_a = v - e_a,
\ee
where $e_a$ is the projection of ${\bf e}_a$ in the $4,5$ plane. We can
define then the manifold $X$ in terms of the $G$ invariants and any
constraints between them. In terms of the invariants
$y=e^{iv'}\prod^d_{a=1}y_a$, $z=e^{-iv'}\prod^d_{a=1}z_a$ and $v$, the
defining equation for the resulting space is given by:
\be
yz = \prod_{a=1}^d (v-e_a),
\ee
which is the equation we have been using in the main text for the
Taub-NUT.

\subsection{Rotating the complex structure}
We have described how to obtain the equations for the Taub-NUT in a
given complex structure, but in our system there are two different
relevant complex structures with no holomorphic relation between them,
and we expect that the $y$, $z$ and $v$ parameters describing the
Taub-NUT in the complex structure in which the NS factor of the M
theory curve is holomorphic have a complicated non-holomorphic relation
with the parameters $y'$, $z'$ and $v'$ describing the curve in the
complex structure where the NS' factor is holomorphic. In this section
we describe how to obtain explicit relations between both sets of
coordinates.

The basic idea has already been described. What we notice is that
rotations of the moment maps have two interpretations, one as rotations
in the space of complex structures of the hyper-K\"ahler manifold and the
other as rotations in the Type IIA theory. Since in the type IIA
theory the rotation necessary for going from the direction associated
with the NS factor being holomorphic (i.e., $x^6$) to the direction in
which the NS' factor is holomorphic ($x^6 \cos \theta + x^4\sin
\theta$, where $\theta$ is the rotation angle, given by the masses and
the position in $x^6$ of the D6 branes) is easy to determine with
simple trigonometry, the appropriate change in complex structure is
simple to determine too. For example, let us identify the $ijk$
directions in quaternion space with the $6,4,5$ directions in the type
IIA picture. With this convention, the complex structure making the NS
factor holomorphic is given by privileging $i$, and splitting the
quaternionic coordinates as $q_a=(a+ib)+j(c+id)=(y_a, z_a)$. Now we
rotate in order to obtain the expressions in the coordinates where NS'
is holomorphic. The effect in $ijk$ is given by:
\be
\left(\begin{array}{c}
i \\ j \\ k
\end{array}\right) \rightarrow \left(\begin{array}{ccc}
\cos \theta & -\sin\theta & 0 \\
\sin\theta & \cos\theta & 0 \\
0 & 0 & 1
\end{array}\right)\left(\begin{array}{c}
i \\ j \\ k
\end{array}\right) \equiv \left(\begin{array}{c}
I \\ J \\ K
\end{array}
\right).
\ee

We now want to split the moment maps into complex coordinates where
$I$ is the privileged complex structure. We just substitute and
read components, let us do it for some generic $q$:
\bea
q  & = & a+bi+cj+dk = a + b(I\cos\theta + J\sin\theta) +
c(J\cos\theta - I\sin\theta) + dk\\
& = & a+I(b\cos\theta - c\sin\theta) + J(b\sin\theta+c\cos\theta+Id).
\eea
From here we read what the new $y$ and $z$ are, and since we have the
expressions of $abcd$ in terms of the original $y$ and $z$ (for
example, $b = -i/2(y-\bar y)$), this completely determines the new
variables as non-holomorphic functions of the old ones and the
$\theta$ angle.


\bibliographystyle{JHEP}

\begin{thebibliography}{99}
\bibitem{Intriligator:2006dd}
  K.~Intriligator, N.~Seiberg and D.~Shih,
  ``Dynamical SUSY breaking in meta-stable vacua,''
  JHEP {\bf 0604}, 021 (2006)
  [arXiv:hep-th/0602239].

\bibitem{Dimopoulos:1997ww}
  S.~Dimopoulos, G.~R.~Dvali, R.~Rattazzi and G.~F.~Giudice,
  ``Dynamical soft terms with unbroken supersymmetry,''
  Nucl.\ Phys.\ B {\bf 510}, 12 (1998)
  [arXiv:hep-ph/9705307].

\bibitem{Franco:2006es}
  S.~Franco and A.~M.~Uranga,
   ``Dynamical SUSY breaking at meta-stable minima from D-branes at 
obstructed
  geometries,''
  JHEP {\bf 0606}, 031 (2006)
  [arXiv:hep-th/0604136].

\bibitem{Berenstein:2005xa}
  D.~Berenstein, C.~P.~Herzog, P.~Ouyang and S.~Pinansky,
  ``Supersymmetry breaking from a Calabi-Yau singularity,''
  JHEP {\bf 0509}, 084 (2005)
  [arXiv:hep-th/0505029].

\bibitem{Franco:2005zu}
  S.~Franco, A.~Hanany, F.~Saad and A.~M.~Uranga,
  ``Fractional branes and dynamical supersymmetry breaking,''
  JHEP {\bf 0601}, 011 (2006)
  [arXiv:hep-th/0505040].

\bibitem{Bertolini:2005di}
  M.~Bertolini, F.~Bigazzi and A.~L.~Cotrone,
  ``Supersymmetry breaking at the end of a cascade of Seiberg
dualities,''
  Phys.\ Rev.\ D {\bf 72}, 061902 (2005)
  [arXiv:hep-th/0505055].

\bibitem{Ooguri:2006pj}
  H.~Ooguri and Y.~Ookouchi,
   ``Landscape of supersymmetry breaking vacua in geometrically realized gauge
  theories,''
  arXiv:hep-th/0606061.

\bibitem{Braun:2006em}
  V.~Braun, E.~I.~Buchbinder and B.~A.~Ovrut,
  ``Dynamical SUSY breaking in heterotic M-theory,''
  arXiv:hep-th/0606166.

\bibitem{Braun:2006da}
  V.~Braun, E.~I.~Buchbinder and B.~A.~Ovrut,
  ``Towards realizing dynamical SUSY breaking in heterotic model building,''
  arXiv:hep-th/0606241.

\bibitem{Garcia-Etxebarria:2006rw}
  I.~Garcia-Etxebarria, F.~Saad and A.~M.~Uranga,
  ``Local models of gauge mediated supersymmetry breaking in string 
theory,''
  arXiv:hep-th/0605166.

\bibitem{Diaconescu:2005pc}
  D.~E.~Diaconescu, B.~Florea, S.~Kachru and P.~Svrcek,
  ``Gauge - mediated supersymmetry breaking in string 
compactifications,''
  JHEP {\bf 0602}, 020 (2006)
  [arXiv:hep-th/0512170].

\bibitem{Elitzur:1997fh}
  S.~Elitzur, A.~Giveon and D.~Kutasov,
  ``Branes and N = 1 duality in string theory,''
  Phys.\ Lett.\ B {\bf 400}, 269 (1997)
  [arXiv:hep-th/9702014].

\bibitem{Giveon:1998sr}
  A.~Giveon and D.~Kutasov,
  ``Brane dynamics and gauge theory,''
  Rev.\ Mod.\ Phys.\  {\bf 71}, 983 (1999)
  [arXiv:hep-th/9802067].

\bibitem{Hori:1997ab}
  K.~Hori, H.~Ooguri and Y.~Oz,
   ``Strong coupling dynamics of four-dimensional N = 1 gauge theories from  M
  theory fivebrane,''
  Adv.\ Theor.\ Math.\ Phys.\  {\bf 1}, 1 (1998)
  [arXiv:hep-th/9706082].

\bibitem{Witten:1997ep}
  E.~Witten,
  ``Branes and the dynamics of {QCD},''
  Nucl.\ Phys.\ B {\bf 507}, 658 (1997)
  [arXiv:hep-th/9706109].

\bibitem{Brandhuber:1997iy}
  A.~Brandhuber, N.~Itzhaki, V.~Kaplunovsky, J.~Sonnenschein and 
S.~Yankielowicz,
  ``Comments on the M theory approach to N = 1 S{QCD} and brane 
dynamics,''
  Phys.\ Lett.\ B {\bf 410}, 27 (1997)
  [arXiv:hep-th/9706127].

\bibitem{deBoer:1998by}
  J.~de Boer, K.~Hori, H.~Ooguri and Y.~Oz,
  ``Branes and dynamical supersymmetry breaking,''
  Nucl.\ Phys.\ B {\bf 522}, 20 (1998)
  [arXiv:hep-th/9801060].

\bibitem{Ooguri:2006bg}
  H.~Ooguri and Y.~Ookouchi,
  ``Meta-Stable Supersymmetry Breaking Vacua on Intersecting Branes,''
  arXiv:hep-th/0607183.

\bibitem{Bena:2006rg}
  I.~Bena, E.~Gorbatov, S.~Hellerman, N.~Seiberg and D.~Shih,
  ``A note on (meta)stable brane configurations in MQCD,''
  arXiv:hep-th/0608157.

\bibitem{Hanany:1996ie}
  A.~Hanany and E.~Witten,
   ``Type IIB superstrings, BPS monopoles, and three-dimensional gauge
  dynamics,''
  Nucl.\ Phys.\ B {\bf 492}, 152 (1997)
  [arXiv:hep-th/9611230].

\bibitem{Barbon:1997zu}
  J.~L.~F.~Barbon,
  ``Rotated branes and N = 1 duality,''
  Phys.\ Lett.\ B {\bf 402}, 59 (1997)
  [arXiv:hep-th/9703051].

\bibitem{Gava:1997jt}
  E.~Gava, K.~S.~Narain and M.~H.~Sarmadi,
  ``On the bound states of p- and (p+2)-branes,''
  Nucl.\ Phys.\ B {\bf 504}, 214 (1997)
  [arXiv:hep-th/9704006].

\bibitem{Witten:1997sc}
  E.~Witten,
  ``Solutions of four-dimensional field theories via M-theory,''
  Nucl.\ Phys.\ B {\bf 500}, 3 (1997)
  [arXiv:hep-th/9703166].

\bibitem{Evans:1997hk}
  N.~J.~Evans, C.~V.~Johnson and A.~D.~Shapere,
  ``Orientifolds, branes, and duality of 4D gauge theories,''
  Nucl.\ Phys.\ B {\bf 505}, 251 (1997)
  [arXiv:hep-th/9703210].

\bibitem{Hori:1998iv}
  K.~Hori,
 ``Consistency condition for fivebrane in M-theory on R**5/Z(2) 
orbifold,''
  Nucl.\ Phys.\ B {\bf 539}, 35 (1999)
  [arXiv:hep-th/9805141].

\bibitem{Gimon:1998be}
  E.~G.~Gimon,
 ``On the M-theory interpretation of orientifold planes,''
  arXiv:hep-th/9806226.

\bibitem{Elitzur:2000pq}
  S.~Elitzur, A.~Giveon, D.~Kutasov, E.~Rabinovici and G.~Sarkissian,
  ``D-branes in the background of NS fivebranes,''
  JHEP {\bf 0008}, 046 (2000)
  [arXiv:hep-th/0005052].

\bibitem{Landsteiner:1998pb}
  K.~Landsteiner, E.~Lopez and D.~A.~Lowe,
  ``Supersymmetric gauge theories from branes and orientifold 
six-planes,''
  JHEP {\bf 9807}, 011 (1998)
  [arXiv:hep-th/9805158].

\bibitem{Landsteiner:1998gh}
  K.~Landsteiner, E.~Lopez and D.~A.~Lowe,
  ``Duality of chiral N = 1 supersymmetric gauge theories via branes,''
  JHEP {\bf 9802}, 007 (1998)
  [arXiv:hep-th/9801002].

\bibitem{Brunner:1998jr}
  I.~Brunner, A.~Hanany, A.~Karch and D.~Lust,
  ``Brane dynamics and chiral non-chiral transitions,''
  Nucl.\ Phys.\ B {\bf 528}, 197 (1998)
  [arXiv:hep-th/9801017].

\bibitem{Elitzur:1998ju}
  S.~Elitzur, A.~Giveon, D.~Kutasov and D.~Tsabar,
  ``Branes, orientifolds and chiral gauge theories,''
  Nucl.\ Phys.\ B {\bf 524}, 251 (1998)
  [arXiv:hep-th/9801020].

\bibitem{Hanany:1997sa}
  A.~Hanany and A.~Zaffaroni,
  ``Chiral symmetry from type IIA branes,''
  Nucl.\ Phys.\ B {\bf 509}, 145 (1998)
  [arXiv:hep-th/9706047].

\bibitem{Brodie:1997sz}
  J.~H.~Brodie and A.~Hanany,
   ``Type IIA superstrings, chiral symmetry, and N = 1 4D gauge theory
  dualities,''
  Nucl.\ Phys.\ B {\bf 506}, 157 (1997)
  [arXiv:hep-th/9704043].

\bibitem{Uranga:1998vf}
  A.~M.~Uranga,
  ``Brane configurations for branes at conifolds,''
  JHEP {\bf 9901}, 022 (1999)
  [arXiv:hep-th/9811004].

\bibitem{Dasgupta:1998su}
  K.~Dasgupta and S.~Mukhi,
  ``Brane constructions, conifolds and M-theory,''
  Nucl.\ Phys.\ B {\bf 551}, 204 (1999)
  [arXiv:hep-th/9811139].

\bibitem{Gibbons:1996nt}
  G.~W.~Gibbons and P.~Rychenkova,
  ``HyperKaehler quotient construction of BPS monopole moduli spaces,''
  Commun.\ Math.\ Phys.\  {\bf 186}, 585 (1997)
  [arXiv:hep-th/9608085].

\end{thebibliography}

\end{document}